\documentclass[aps,prl,superscriptaddress,twocolumn,10pt]{revtex4-2}

\usepackage[utf8]{inputenc}
\usepackage[version=3]{mhchem}
\usepackage{amsmath}
\usepackage{amsfonts}
\usepackage{hyperref}
\usepackage{braket}
\usepackage{lipsum}
\usepackage{url}
\usepackage[draft]{graphicx}
\usepackage{lipsum}
\usepackage{color}
\usepackage{float}
\usepackage[export]{adjustbox}
\usepackage{wrapfig}
\usepackage{array,multirow}
\usepackage{tabularx}
\usepackage{bm}

\usepackage{relsize}  

\makeatletter
\renewcommand\frontmatter@abstractwidth{\dimexpr\textwidth-1in\relax}
\makeatother
\usepackage[normalem]{ulem}
\usepackage{color}
\usepackage{cancel}
\definecolor{my_red}{rgb}{.7,0,0}
\definecolor{my_green}{rgb}{0,.7,0}
\definecolor{my_blue}{rgb}{0,0,.7}
\definecolor{my_purple}{rgb}{.7,0,.7}

\begin{abstract}
We show that applying a few-femtosecond mid-infrared laser pulse parallel to the backbone of a halogenated, conjugated organic molecule induces localized ionization followed by long-lasting field-free charge migration (CM). Through time-dependent density-functional theory calculations, we find that this CM is robust with respect to the parameters of the applied laser pulse. We document the spatial and temporal origin of the localized hole, which is initiated via strong-field tunnel ionization in the halogen end of the molecule, and near the peak of the laser electric field. In bromobutadiyne we find a range of wavelengths around $1500~\text{nm}$ that consistently leads to the initiation of high-contrast CM. In addition, we show that the inclusion of Ehrenfest nuclear dynamics does not disrupt the creation of the localized hole, nor the subsequent CM motion, in bromobutadiyne or \emph{para}-bromoaniline.
\end{abstract}

\begin{document}

\author{Kyle~A.~Hamer}
\email{khamer3@lsu.edu}
\affiliation{Department of Physics and Astronomy, Louisiana State University, Baton Rouge, LA 70803, USA}

\author{Fran\c{c}ois~Mauger}
\affiliation{Department of Physics and Astronomy, Louisiana State University, Baton Rouge, LA 70803, USA}

\author{Kenneth~Lopata}
\affiliation{Department of Chemistry, Louisiana State University, Baton Rouge, LA 70803, USA}
\affiliation{Center for Computation and Technology, Louisiana State University, Baton Rouge, LA 70803, USA}

\author{Kenneth~J.~Schafer}
\affiliation{Department of Physics and Astronomy, Louisiana State University, Baton Rouge, LA 70803, USA}

\author{Mette~B.~Gaarde}
\email{mgaarde1@lsu.edu}
\affiliation{Department of Physics and Astronomy, Louisiana State University, Baton Rouge, LA 70803, USA}

\title{Strong-field ionization with few-cycle, mid-infrared laser pulses induces a localized ionization followed by long-lasting charge migration in halogenated organic molecules}
\date{\today}
\maketitle

Charge migration (CM), the coherent motion of an electron hole in a molecule following a sudden ionization or excitation event, is a topic that has attracted much attention within the ultrafast community \cite{weinkauf1997, cederbaum1999, breidbach2003, calegari2014, kraus2015, worner2017, nisoli2017}. 
The electronic nature of CM means that it typically occurs on attosecond timescales, and the study of CM is critical for understanding, and perhaps steering, downstream processes such as photosynthesis, photocatalysis, and light harvesting \cite{dutoi2011, eberhard2008, remacle2006}. First discussed in theoretical studies in the late 1990s \cite{weinkauf1997, cederbaum1999, breidbach2003}, several experimental measurements of CM in various classes of molecules have been reported in the last decade \cite{calegari2014, calegari2016, worner2017, nisoli2017, grell2023, he2023}. 

Experiments probing CM have generally initiated the electron hole in one of two ways: by one-photon ionization or excitation using an attosecond extreme ultraviolet (XUV) or soft X-ray pulse \cite{calegari2014, barillot2022, grell2023}, or by tunnel ionization using an infrared laser pulse \cite{kraus2015, matselyukh2022, he2022}. 
Although early theoretical studies of CM often used a hole initiated by the \emph{ad hoc} removal of an electron from some orbital \cite{hennig2005, despre2019, despre2022, chordiya2023, jia2017}, recent works have used a superposition of initially-populated cationic states similar to what would be created from one-photon ionization \cite{perfetto2018, kuleff2016, lara-astiaso2018, khalili2021}. Several works have shown that tunnel ionization can induce ultrafast electron dynamics \cite{smirnova2009, kraus2015, schlegel2022}, including CM in simple two-center molecules \cite{kraus2015}, but the dynamics of the hole-formation process and the characteristics of the hole it leaves behind have been largely unexplored. The character of the initial hole has been shown to be important for the subsequent charge migration: several papers have demonstrated that a higher degree of spatial localization of the initial hole leads to higher-contrast, particle-like CM dynamics that is robust against small changes in the initial condition and thus a desirable candidate for experimental study~\cite{folorunso2021, mauger2022, hamer2022, folorunso2023, hamer2023}. 

In this Letter, we show that a spatially localized hole can be created via strong-field tunnel ionization by a mid-infrared (MIR) few-cycle laser pulse. We demonstrate this in two different molecules: bromobutadiyne (BrC$_{4}$H) and \emph{para}-bromoaniline ($p$-BrC$_{6}$H$_{4}$NH$_{2}$). Specifically, using time-dependent density-functional theory (TDDFT) \cite{runge1984, gross2012}, we show that a few-cycle pulse polarized along the backbone of a halogenated, conjugated organic molecule creates a localized hole on the halogen atom via strong-field tunnel ionization, followed by sustained CM motion. During the laser pulse, the electron density predominantly ionizes near the extrema of the electric field from the bromine end of the molecule, creating a localized hole. After the laser pulse has ended, this localized hole travels freely along the molecular backbone, via the molecule's $\pi$-bond structure, with a different frequency than that of the ionizing laser. In bromobutadiyne, we find that this CM mode is robust with respect to the laser pulse parameters, as long as the ionization process happens primarily within one half-cycle of the laser and is dominated by tunnel ionization. Finally, we show that tunnel ionization induces a qualitatively-similar CM mode in \emph{para}-bromoaniline, and that the inclusion of Ehrenfest nuclear dynamics for a single, most probable, configuration in both molecules does not disrupt the creation of the localized hole nor its subsequent high-contrast CM motion.

We perform our calculations with grid-based TDDFT using a local-density-approximation (LDA) exchange-correlation functional~\cite{perdew1981, yabana1996, marques2012} and an average-density self-interaction correction (ADSIC)~\cite{ciofini2003, tsuneda2014} within the Octopus software package~\cite{marques2003, andrade2015, tancogne-dejean2020}. We use a simulation box with dimensions of $40 \times 40 \times 90$ atomic units (a.u.), with the longer dimension parallel to the molecular backbone, and a grid spacing of 0.3~a.u.\ in all directions.
To remove ionized electron densities, we use a $\sin^{2}$ complex absorbing potential that extends 12.5 a.u.\ from the edges of the box. 

\begin{figure}[tb]
    \centering
    \includegraphics[width=\linewidth]{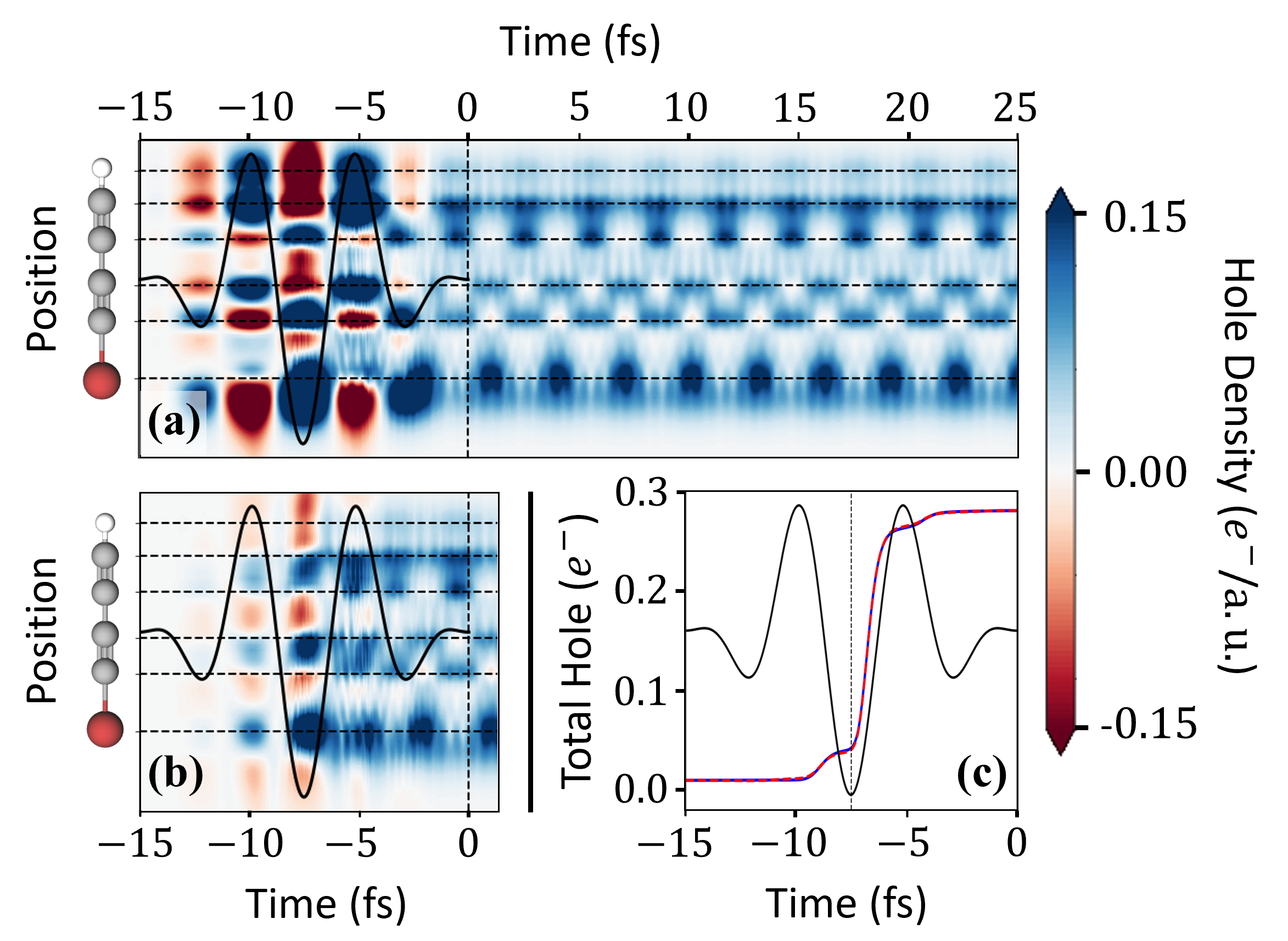}
    \caption{(a) Time-dependent hole density in bromobutadiyne during (negative times) and after (positive) a 3-cycle laser pulse polarized parallel to the molecular backbone. Here we show the hole density integrated along the two dimensions perpendicular to the molecular backbone. (b) Same as (a), but with the effect of the first-order laser-induced polarization removed (see text). (c) Time-dependent ionization yield during the laser pulse for the unfiltered (solid blue curve) and polarization-filtered (dashed red) dynamics of panels (a) and (b), respectively. In all panels, the thin black curve sketches the pulse electric field, with 1500-nm wavelength and 50-TW/cm$^{2}$ peak intensity.}
    \label{fig:1}
\end{figure}

Figure~\ref{fig:1}~(a) illustrates the strong-field-induced CM dynamics in the bromobutadiyne molecule during and after a few-laser-cycle ionizing pulse.
Specifically, we show the time-dependent hole density $\rho_{h}(z, t)$, defined as the difference between the neutral ground-state density and the time-dependent cationic density, integrated over the directions transverse to the molecular backbone.
We set the time $t = 0$ such that the laser driven dynamics takes place for $t < 0$ and the field-free dynamics for $t > 0$. Here, the molecule starts in its neutral ground state, and then we apply a 1500-nm ($\omega_{L} = 0.83$~eV), 50-TW/cm$^{2}$ laser pulse polarized along the molecular backbone. All the results we show in this Letter use a total pulse duration of 3 optical cycles, and the carrier-envelope phase (CEP) of the laser pulse is chosen such that the global extremum of the electric field points towards the bromine, to promote ionization from that end of the molecule -- see the black curves in the figure. 
During the laser pulse, the molecule is strongly polarized, such that the electron density is strongly driven by the Lorentz force. Once the pulse has ended, the dynamics settles into a long-lasting, high-contrast CM mode with a period of 3.0~fs ($\omega_{\text{CM}} \approx 1.4\ \text{eV}$). 
With our choice of laser parameters, in total the applied laser pulse ionizes approximately 0.3 electrons, most of it happening near the global extremum of the field at $t = -7.5\ \text{fs}$, as shown in Fig.~\ref{fig:1}~(c). 

As can be seen in Fig.~{\ref{fig:1}}~(a), because of the large polarizability of the target, the valence electron density is strongly distorted by the applied laser field. This manifests as large variations in the electron density around the two ends of the molecule, with a reduction in the density in the direction of the laser electric field and an increase in the opposite direction -- respectively positive and negative hole densities in the plot.
For our purpose, the strong polarizability obscures the direct visualization of the hole being created by the strong-field ionization. To reveal the hole creation, 
we filter out the first-order polarizability by subtracting the part of the time-dependent hole density that is proportional to the electric field:
\begin{equation}\label{eq:pol_filt}
    \tilde{\rho}_{h}(z, t) = \rho_{h}(z, t) - \alpha(z) \cdot E_{z}(t).
\end{equation}
We determine $\alpha(z)$ by minimizing $\tilde{\rho}_{h}(z,\ t)$ for the first two femtoseconds of the pulse, before nonlinear effects kick in, using a least-squares fitting at every value of $z$ and show the result in Fig.~{\ref{fig:1}}~(b). Note that, by definition, the original and polarization-filtered dynamics are identical when the field is off, for $t \geq 0$. 
We have also checked that the polarization filter preserves the time-dependent ionization yield, as evidenced by the almost-perfect match between the original (solid blue curve) and filtered (dashed red) yields in Fig.~\ref{fig:1}~(c). Panel (c) also reveals that ionization primarily happens shortly after the largest peak of the laser pulse. The short delay between the global extremum of the field (at $t=-7.5$~fs) and the steep jump in the ionization yield is due to the time it takes for the ionized density to reach the absorbing boundary, where it is removed from the simulation box. Fig.~{\ref{fig:1}}~(b) shows that the ionization-induced hole predominantly forms around the bromine atom. After its creation, the laser field drives this hole up and down the molecular backbone at the frequency of the laser, as evidenced by the valence hole on the terminal bond near $t = -5~\text{fs}$. The laser field thus preserves the hole's localized nature until the field-free dynamics takes over near $t = 0$. 

\begin{figure}[tb]
    \centering
    \includegraphics[width=\linewidth]{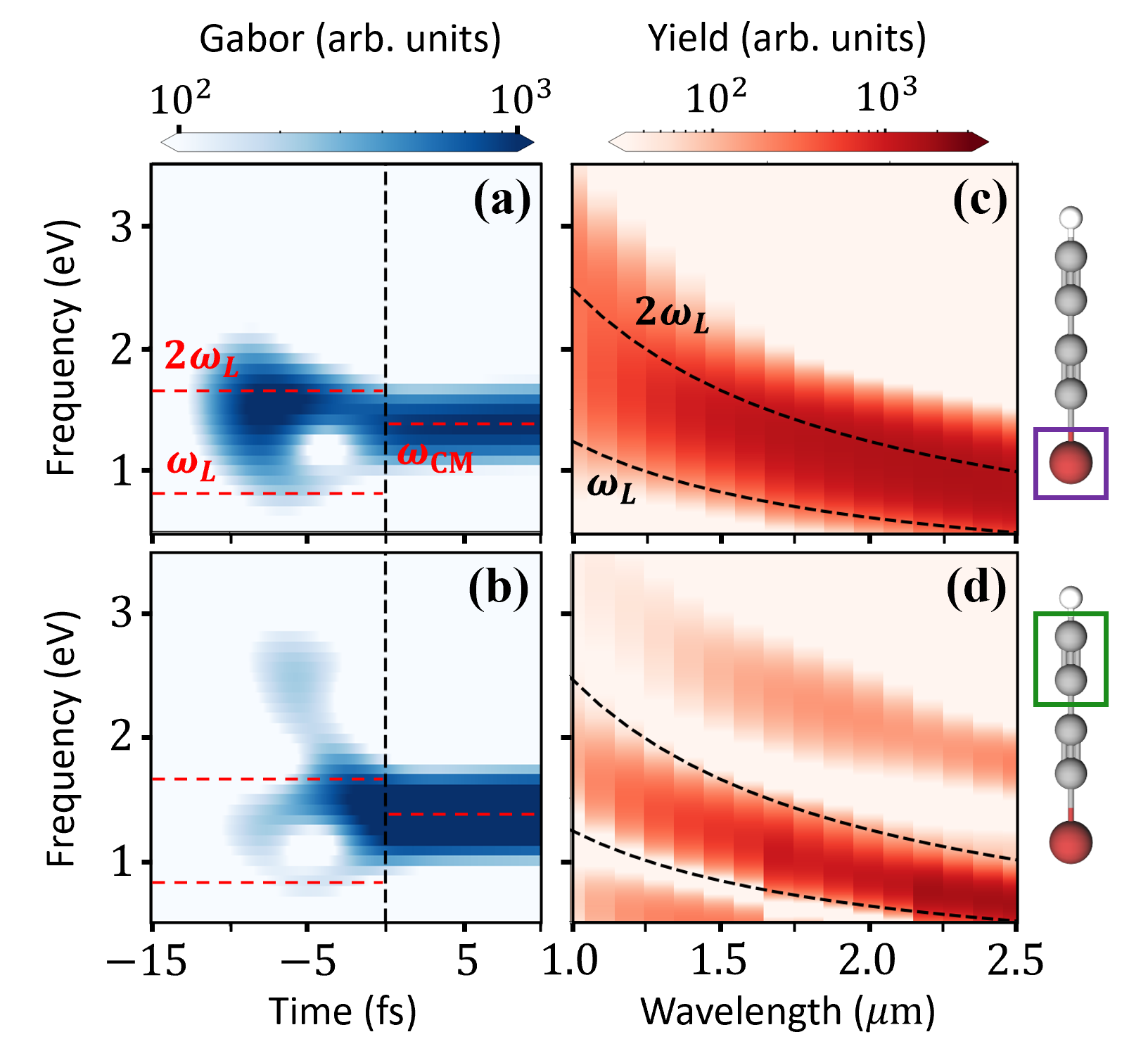}
    \caption{(a, b) Gabor transforms of the polarization-filtered hole dynamics of Fig.~\ref{fig:1}~(b) around the (a) bromine and (b) terminal bond ends of the molecule. The horizontal dashed lines mark the laser and CM frequencies during and after the pulse, respectively. The width of the window used in the Gabor transform is approximately 18.7~fs. (c, d) Wavelength dependence of the polarization-filtered hole dynamics around the same two ends of the molecule. For each wavelength, the colormap is obtained by taking the intensity of Fourier transform of the polarization-filtered hole dynamics over the entire duration of the driving field, fixed at 3 laser cycles and same peak intensity and CEP as in Fig.~\ref{fig:1}.}
    \label{fig:2}
\end{figure}

It is clear from Fig.~{\ref{fig:1}} that the frequencies of the laser-driven and the field-free hole dynamics are different. To explore this in more detail, we next perform a time-frequency analysis of the hole density around the two ends of the molecule and show the results in Fig.~\ref{fig:2}. Specifically, we integrate the polarization-filtered time-dependent hole density in small regions around the bromine and terminal C$\equiv$C bond (1 and 3~a.u.\ width, respectively-- see the sketched on the right of the figure), and perform a Gabor transform of both time-dependent signals. 

Figure~\ref{fig:2}~(a) shows that, during the laser pulse, the polarization-filtered hole dynamics on the bromine end of the molecule is periodic at twice the laser frequency: the spectrogram is centered around the upper dashed horizontal line at $2\omega_{L} = 1.67$~eV for $t<0$. This is in contrast to the terminal bond polarization-filtered spectrogram of Fig.~\ref{fig:2}~(b), which does not exhibit any prominent periodicity during the laser pulse. We note that the polarization filtering removes a strong $\omega_{L}$ periodicity at both ends of the molecule, but the polarization response does not contribute to the ionization yield -- see Fig.~\ref{fig:1}~(c). After the laser pulse is over, the spectrograms at both ends of the molecule switches to the CM-mode frequency $\omega_\text{CM}\approx 1.4$~eV, with a small contribution from the $2\omega_\text{CM}$ harmonic that is characteristic of CM modes~\cite{folorunso2021, hamer2022}. The transition between the laser-driven and field-free frequencies is determined by the width of the window used in the Gabor transform, which must be larger than both the laser and CM periods. Altogether, the results of Fig.~{\ref{fig:2}}~(a,b) further support our observation from Fig.~{\ref{fig:1}}~(b) that the strong-field tunnel ionization favors the bromine end of the molecule, in agreement with Ref.~\cite{kraus2015}.

Next, we investigate the effect of the driving wavelength on the strong-field tunnel ionization process and the field-free CM it induces. For this, we fix the pulse duration to 3 optical cycles total, the peak intensity to 50 TW/cm$^{2}$, and the CEP while varying the laser wavelength. Figures~\ref{fig:2}~(c,d) show the evolution of the Fourier transform of the polarization-filtered hole density, over the entire duration of the pulse and around the two ends of the molecule as functions of the wavelength. Irrespective of the wavelength, from 1.0 to 2.5~$\mu$m, the bromine end of the molecule again exhibits a strong $2\omega_{L}$ feature, which is absent at the terminal-bond end of the molecule -- compare panels (c) and (d), respectively.
This demonstrates that MIR strong-field ionization robustly induces a localized hole on the halogen end of the target \cite{sandor2019, schlegel2022}.

\begin{figure}[tb]
    \centering
    \includegraphics[width=\linewidth]{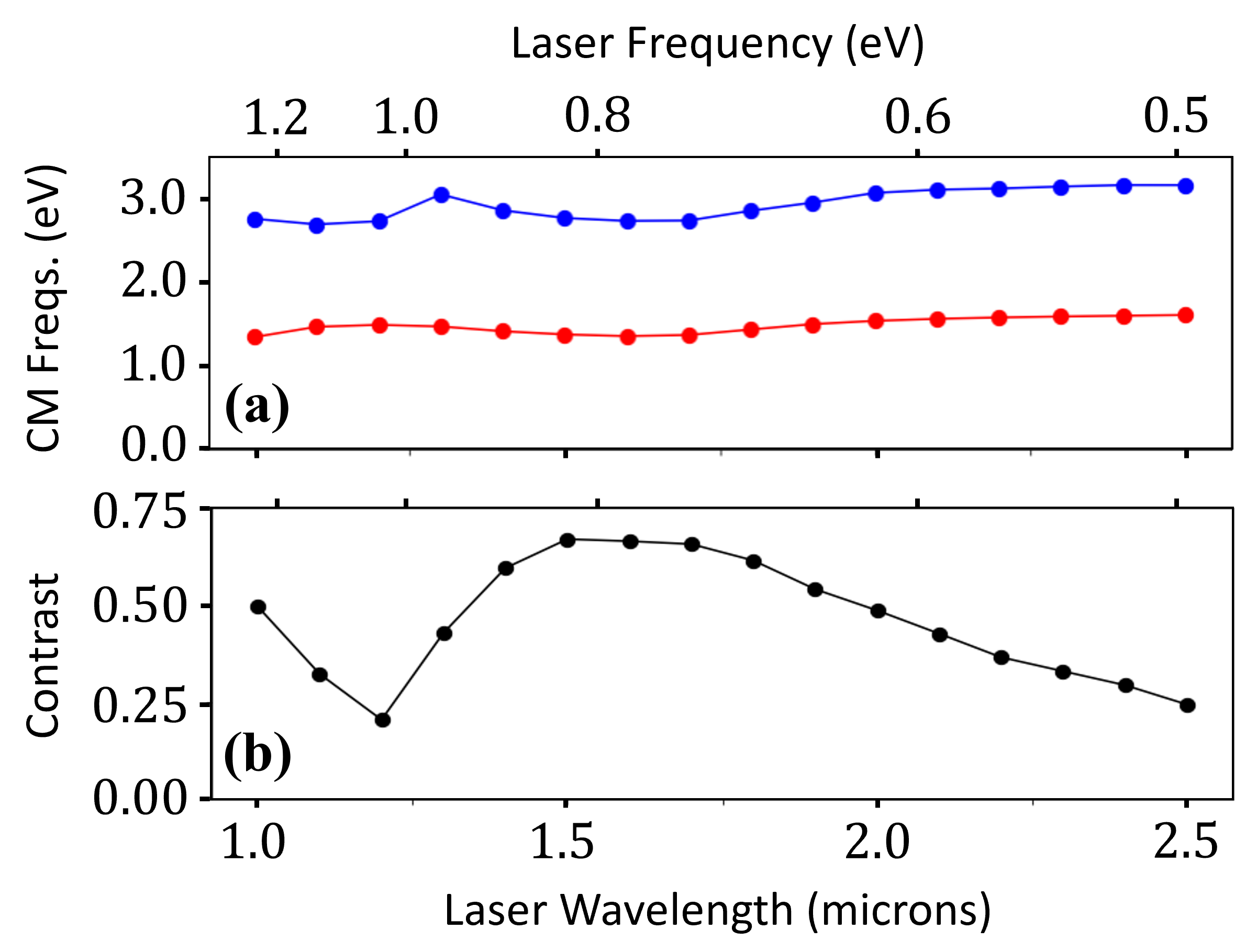}
    \caption{Field-free CM (a) frequencies and (b) contrast (see text) induced by a 3-optical cycles, 50 TW/cm$^{2}$ pulse, as functions of the laser wavelength.}
    \label{fig:3}
\end{figure}

We now turn to the field-free CM and first consider the effect of the ionizing-pulse wavelength on the CM frequency. Specifically, we obtain the CM frequencies by constructing a complex signal $\chi(t)$ such that $\text{Re}[\chi(t)]$ is the hole density integrated around the bromine atom and $\text{Im}[\chi(t)]$ is the hole density integrated around the terminal C$\equiv$C bond. We then detect the two strongest non-zero frequency components of $\chi(t)$~\cite{mauger2022} and show the results in Figure~\ref{fig:3}~(a). Strikingly, we see that the fundamental field-free CM frequency (red curve) is robust in a range of wavelengths between 1.0 and 2.5~$\mu$m at $\omega_\text{CM} \approx 1.4$~eV. We also see that the second leading frequency (blue curve) is locked at $2\omega_{\text{CM}}$ everywhere the field-free CM has high contrast, which is a hallmark of a robust CM mode~\cite{hamer2022,mauger2022}.
Combined with the results of Fig.~\ref{fig:2}~(c,d), we see that the CM-creation process has two intrinsic time scales. The first one is imposed by the laser pulse, during the ionization process and the creation of a localized hole on the halogen end of the molecule. The second one is imposed by the target, after the end of the pulse, leading to a consistent field-free CM frequency over a wide range of MIR ionization wavelengths.

To quantify the overall efficacy of the CM-creation process, we calculate the CM dynamics contrast as a function of the laser wavelength. To calculate the contrast, we first fit the hole density integrated around the bromine atom to $A\cos{\left(\omega_{\text{CM}} t + \phi \right)} + B$, where we get $\omega_\text{CM}$ from the frequency analysis of Fig.~\ref{fig:2}~(a). Then we define the CM contrast as the ratio $c_\text{CM}=A/B$ and show the result in Fig.~\ref{fig:2}~(b); by construction, $0 \leq c_\text{CM} \leq 1$. In our case, we clearly see a range of wavelengths near $1500$~nm where the induced CM mode in bromobutadiyne is robust and has a high contrast. Above 1800 nm, tunnel ionization heavily favors removing charge from the highest-occupied molecular orbital. This puts the molecule close to its cationic ground state, thereby lowering the contrast of the CM. Therefore, we expect that some amount of non-adiabaticity is needed in the ionization process to sustain the spatially-localized hole required for field-free CM. 
Around 1200~nm, we observe a dip in the contrast, likely due to some target-specific resonant process around this wavelength.

Overall, the results of Fig.~{\ref{fig:3}} show that few-cycle MIR strong-field ionization in bromobutadiyne creates an electron hole with some degree of spatial localization around the halogen, followed by a robust CM mode with a frequency of 1.4~eV \cite{mauger2022}. In addition to the ionizing pulse wavelength, we have checked that the induced CM mode in bromobutadiyne is robust with respect to the laser's intensity, pulse duration, and CEP, as well as the angle between the polarization axis and the molecular backbone (not shown). Specifically, while the contrast varies with the laser parameters and molecular orientation, we systematically observe the same 1.4-eV CM mode.

\begin{figure}[tb]
    \centering
    \includegraphics[width=\linewidth]{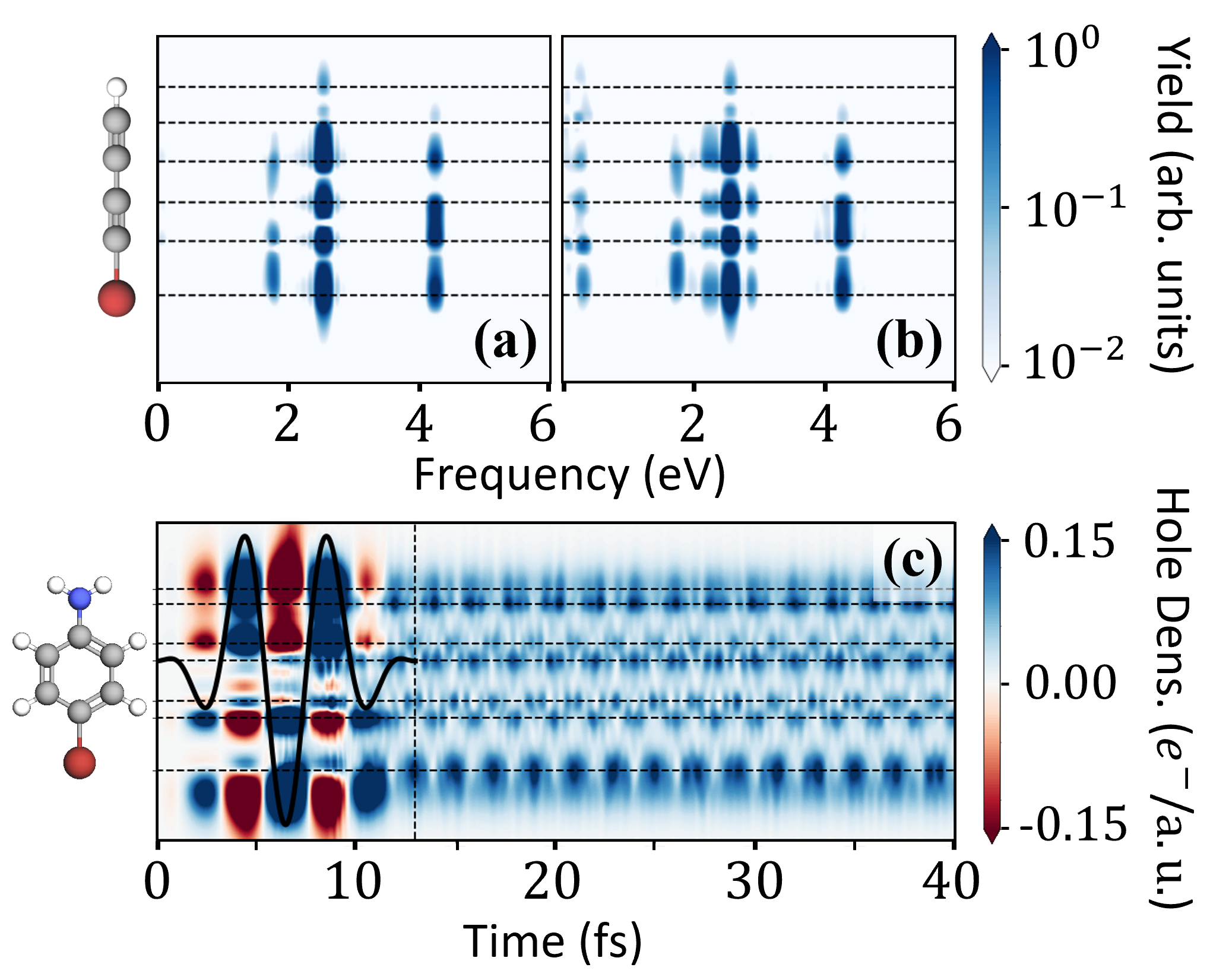}
    \caption{Fourier transform of the field-free hole dynamics in bromobutadiyne induced by a 3-cycle, 1000-nm, 50-TW/cm$^2$, pulse (a) without and (b) with Ehrenfest nuclear dynamics. Here we start our simulations from an optimized geometry leading to a slightly shorter molecule than in the previous figures (see text). (c) Time-dependent hole density in \emph{para}-bromoaniline during and after a 3-cycle, 1300-nm, 25-TW/cm$^2$ pulse.}
    \label{fig:4}
\end{figure}

Finally, we investigate the extent to which nuclear dynamics may influence the robustness of the hole creation process and its subsequent CM dynamics. We do this by including nuclear dynamics, at the Ehrenfest level, into our TDDFT simulations. The first step in these simulations is to optimize the molecular geometry for the specific DFT functionals we use in the time propagation. In our case it leads to a shrinkage of the bromobutadiyne molecular model by about 1~\AA\ as compared to the more realistic configuration we have used above~\cite{folorunso2021, hamer2022}. The shorter molecule  leads to a higher mode frequency without otherwise qualitatively affecting the overall dynamics. Starting from the optimized neutral geometry, Figs.~{\ref{fig:4}}~(a,b) compare the Fourier transform of the field-free hole dynamics in bromobutadiyne (a) without and (b) with the inclusion of Ehrenfest nuclear dynamics following strong-field ionization by a 3-cycle, 1000-nm, 50-TW/cm$^2$ pulse.

As was done in Ref.~{\cite{yu2023b}}, to make a meaningful comparison between Figs.~{\ref{fig:4}}~(a) and~(b), we redefine the time-dependent hole density to account for nuclear motion as
\begin{equation}\label{eq:hole_dens_mod}
    \rho_{h}(\vec{r}, t) = \rho[\{\vec{R}(t)\}](\vec{r}, t) - \rho_\circ[\{\vec{R}(t)\}](\vec{r})
\end{equation}
where $\rho[\{\vec{R}(t)\}](\vec{r}, t)$ is the time-dependent electron density from the Ehrenfest TDDFT calculation, and $\rho_\circ[\{\vec{R}(t)\}](\vec{r})$ is the ground-state electron density of the neutral calculated for the time-dependent nuclear geometry $\{\vec{R}(t)\}$. This redefinition of the hole density removes low-frequency contributions which occur due to the motion of the electron density that is localized around each of the moving atomic centers.

Strikingly, in Fig.~\ref{fig:4} all the purely-electronic features of panel (a) are repeated when the Ehrenfest nuclear dynamics is turned on in (b). In the latter, nuclear vibration manifests as a low-frequency feature around 0.2~eV and two sidebands at 2.5$\pm$0.2~eV around the CM frequency. This type of sidebands was also observed in the calculations of Ref.~{\cite{yu2023b}}. The time-dependent hole densities (not shown) corresponding to Figs.~{\ref{fig:4}}(a) and~(b) are qualitatively very similar, since the nuclear dynamics is relatively weak. Thus, we conclude that the inclusion of Ehrenfest nuclear dynamics does not change either the initiation nor the propagation of the CM mode. A thorough investigation into the effect of including nuclear degrees of freedom in bromobutadiyne, including sampling over different initial geometries, can be found in Ref.~{\cite{yu2023b}}. In short, the authors found that though each nuclear configuration yielded long-lasting CM dynamics, the frequency of this dynamics varies by up to 0.25~eV depending on the nuclear geometry. Thus, we expect that decoherence of the CM dynamics could arise from the coherent averaging of many molecules undergoing CM at slightly different geometries. 

As a further demonstration of the robustness of few-cycle MIR tunnel ionization in inducing field-free CM dynamics, in Fig.~{\ref{fig:4}}~(c) we show the time-dependent hole density of Eq.~\eqref{eq:hole_dens_mod} for a \emph{para}-bromoaniline molecule irradiated by a 3-cycle, 1300-nm, 25-TW/cm$^2$ laser pulse, including Ehrenfest nuclear dynamics. Compared to bromobutadiyne, we adjust the field intensity and wavelength to account for \emph{para}-bromoaniline's lower ionization potential and optimal CM-inducing laser configuration. During the laser pulse ($t<0$), the electron density is strongly affected by the Lorentz force, as was also seen in Fig.~{\ref{fig:1}}~(a). After the pulse ($t>0$), the hole dynamics settles into a long-lasting CM mode, which spans the entire length of the molecule and has a period of 2.0~fs ($\omega_{\text{CM}} \approx 2.0$~eV). When removing the Ehrenfest nuclear dynamics, we find that the hole density in both the time and frequency domains looks identical to Fig.~{\ref{fig:4}}~(c) and its Fourier transform, respectively. 

In conclusion, we have shown that strong-field tunnel ionization with a few-cycle MIR laser pulse can create a localized electron hole, both in time and space, on the halogen end of a halogenated, conjugated organic molecule. This localized ionization then induces a robust CM mode along the entire backbone of the molecule after the laser pulse has ended. The frequency of the CM mode is independent of the ionizing-laser wavelength, as shown in Fig.~\ref{fig:3}~(a), and also robust with respect to the pulse duration, CEP, and angle between the molecular backbone and the electric field. Finally, we have found that the inclusion of nuclear dynamics via Ehrenfest TDDFT does not qualitatively affect the CM mode. 

Given that tunnel ionization generates CM mode dynamics in two molecules as different as bromobutadiyne and \emph{para}-bromoaniline, we expect that our results will work for other functionalized halobenzenes and perhaps bio-relevant molecules as well. Our preliminary results show that multiphoton ionization (0.5 - 1.0 $\mu$m) also induces CM dynamics, with a lower contrast but with the same CM frequency. Also, we find that one-photon ionization out of localized inner-valence orbitals, using a 20-50 eV attosecond pulse, does not induce the same type of particle-like, high-contrast CM dynamics as shown above. Further investigations into the shorter-wavelength regime could provide interesting insights into these observations.


{\bf Data availability}: The TDDFT simulation data and Python scripts we use to produce the figures are available at [authors will provide link to public repository for production].


\begin{acknowledgments}

This work was supported by the U.S.\ Department of Energy, Office of Science, Office of Basic Energy Sciences, under Award No.~DE-SC0012462.
Portions of this research were conducted with high performance computational resources provided by Louisiana State University (\url{http://www.hpc.lsu.edu}) and the Louisiana Optical Network Infrastructure (\url{http://www.loni.org}).
\end{acknowledgments}


\end{document}